\documentclass[slac_one]{revtex4}
\usepackage{graphicx}
\usepackage{fancyhdr}
\begin{document}

\title{The PANDA Detector at FAIR} 

%

\author{S.~Marcello (for the PANDA Collaboration)}
\affiliation{Dipartimento di Fisica Sperimentale, Universit\`a 
di Torino and INFN, via P. Giuria 1, 10125, Torino, Italy}

\begin{abstract}
The PANDA detector is under design to be installed at the
HESR storage ring for antiproton of the future FAIR facility in Darmstadt, 
Germany. Fundamental questions of hadron and nuclear physics
interactions of antiprotons with nucleons and nuclei will be pursued using 
a multipurpose set-up which includes innovative detectors. 
Here, the FAIR facility and the PANDA detector are described. 
\end{abstract}

\maketitle

\thispagestyle{fancy}


\section{PANDA EXPERIMENT AT FAIR} 

The PANDA spectrometer~\cite{panda-TPR} is one of the large apparatuses 
at the future Facility for Antiproton and Ion Research (FAIR)
in Europe. Gluonic excitations and the physics of strange and charm 
quarks will be accessible with unprecedented accuracy, allowing 
high-precision tests of strong interaction. 

\begin{figure*}[t]
\centering
\includegraphics[width=100mm]{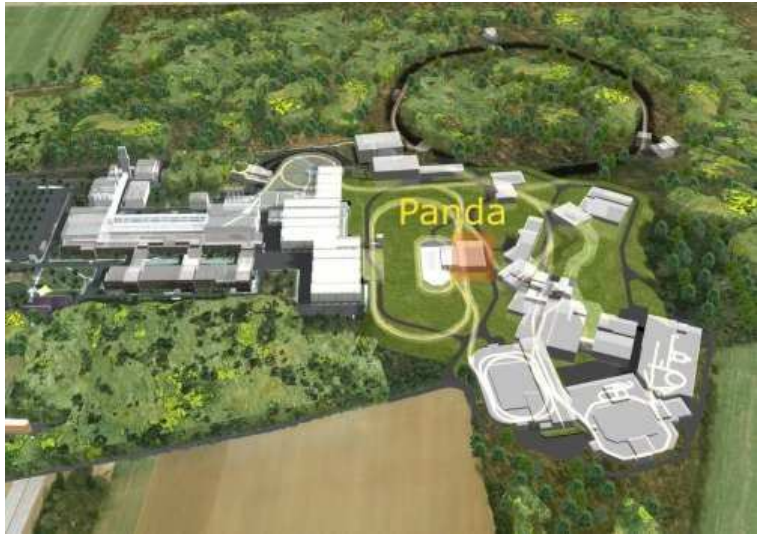}
\caption{Aerial view of the future FAIR facility with the HESR 
(yellow coulor) and the PANDA hall (orange coulor).} \label{fair}
\end{figure*}

The FAIR facility (shown in Figure~\ref{fair}) is under construction
at the GSI Laboratory in Germany. Science goals of the international
FAIR project span a broad range of research fields, from atomic and plasma
physics to the structure of the matter, from the quark-gluon
structure of hadrons to physics of astronomical objects.
The existing linear and SIS18 accelerators at GSI will be used as injector for 
the new system. The hearth of FAIR is a double ring accelerator SIS100/300
of 1~km diameter, which will provide primary beams of $^{238}$U ions in a 
high charge state,
at an energy up to 35~AGeV, and of protons up to 30~GeV. The beams will be sent
to several branches to be used by different experiments.
Secondary beams of radioactive ions of 2~AGeV energy and 
$\bar{p}$ up to a momentum of 15~GeV/c will be available.
The $\bar{p}$ will be pre-cooled in a system of storage and cooler 
rings (RESR/CR)
and then injected at 3.7~GeV/c in the High Energy Storage Ring (HESR),
where the PANDA detector will be installed.
Commissioning of $\bar{p}$ beam for PANDA is forseen in 2015.
The HESR is a storage ring for $\bar{p}$ in the momentum range from 
1.5 to 15~GeV/c. An important feature of HESR is the combination of 
phase-space cooled beams and internal target, 
which will allow experiments of unprecedented precision.
Two different operation modes will be available at HESR: high resolution mode
using the electron cooling, allowing to achieve a very high momentum 
resolution $\delta p/p \sim 10^{-5}$ at a moderate luminosity of 
$10^{31} cm^{-2}s^{-1}$, 
and high luminosity mode using the stochastic cooling, 
with greater momentum spread ($\delta p/p \sim 10^{-4}$),
but at $\cal{L}$ $= 2 \times 10^{32} cm^{-2}s^{-1}$.

\section{THE PHYSICS PROGRAMME}

At HESR the combination of a cooled beam and an internal target 
will allow to study
charmonium spectroscopy with unprecedented precision.
Indeed, the precision of the resonance mass and width measurements  
depends on the precision of the beam energy and 
beam energy width measurements, respectively. The determination of both these 
quantities is based on the measurement of the beam revolution frequency
spectrum.
In the PANDA experiment the detector is used 
to identify and reconstruct
the final state and not to measure the parameters of the resonance. 
Thus the experimental
detector resolution determines the sensitivity to a given final state.
Since the $\bar{p}$  can be cooled very effectively an excellent resonance
mass resolution of 30~keV can be achieved, which has to be compared with
previous best values obtained in $\bar{p}-p$ annihilations 
(100~keV) or in $e^+ - e^-$ annihilations (10~MeV) . 

Studies of hadron structure can be performed with different probes,
each one with its specific advantages.
The reason to use $\bar{p}-p$ annihilations instead of $e^+ - e^-$ ones
is that all states can be directly formed and the mass
resolution is very good.
On the contrary, in $e^+ - e^-$ interactions only states with the 
quantum numbers of the photon,
J$^{PC}$=$1^{--}$, can be formed, all other states being
accessible by secondary decays with limited statistics, thus
a moderate mass resolution can be achieved. 
Moreover, $\bar{p}-p$ gives access to Boer-Mulders function (with no polarised
beam) and to transversity distribution.

The PANDA Physics programme~\cite{panda-TPR} concerns not only
charmonium spectroscopy but spans many different topics, 
all of them addressed 
to answer to fundamental questions of hadron and nuclear physics.
Search for gluonic excitations, such as glueballs and hybrids, will be pursued
in the charmonium region. 
Properties of mesons with hidden and open charm in the nuclear
medium will be studied to understand the origin of hadron masses.
Moreover, the open charm physics will be studied, measuring
the rich spectrum of D mesons and their dominant and rare decays.
Since in PANDA the $\Xi$ and $\bar{\Xi}$ hyperons 
will be copiously produced,
it will be possible to generate double-$\Lambda$ hypernuclei, which allow
to study the Y-N and Y-Y interactions.
Last, but not least, the study of nucleon structure will be investigated,
by measuring Generalized Parton Distributions (Drell-Yan and Deeply
Virtual Compton Scattering, "spin" structure functions by polarised 
$\bar{p}$ and proton time-like form factors up to
25~GeV c.m. energy (never reached before).

\section{THE DETECTOR}

Such an ambitious programme requires a high performance spectrometer. 
The main requirements can be summarised as follows:
\begin{itemize}
\item Capability to handle an interaction rate up to 20~MHz, 
especially for detectors closed to the target and in the forward direction;
\item Full angular coverage for Partial Wave Analysis;
\item Momentum resolution of 1$\%$;
\item High vertex resolution $< 100 \mu$m to reconstruct secondary vertex 
of D meson; 
\item Good PID for charged particles in a large range of energies;
\item Efficient $\gamma$ detection in a large range of energies 
(10~MeV$\div$10~GeV);
\item Efficient event selection.
\end{itemize}

\begin{figure*}[t]
\centering
\includegraphics[width=135mm]{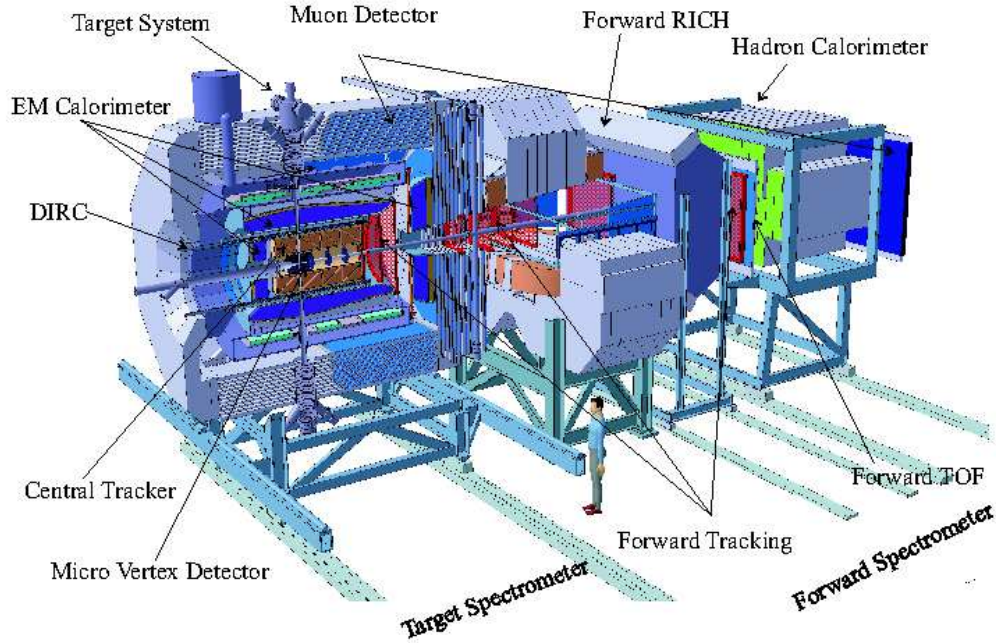}
\caption{3D CAD overview of the PANDA detector. Target Spectrometer 
and Forward Spectrometer with tracking detectors, EM and hadron calorimeters 
and $\check{C}$erenkov detectors.} \label{panda_det}
\end{figure*}

The experimental set-up is shown in Figure~\ref{panda_det}.
The apparatus is a combination of a target and a forward spectrometer
(TS and FS in the following)
of modular design, optimized for the specific kinematics of the 
$\bar{p}-p$ annihilation process with energy range 1.5$\div$15~GeV/c.
The basic concept of the TS is a shell-like arrangement 
of various detector systems surrounding the interaction point of $\bar{p}$
in a fixed target inside a magnetic field
of 2~T provided by a large solenoid. To overcome the gap in
acceptance and resolution in the forward region of such a system 
a FS with a dipole magnet of large gap (1~m) will be used.
It will cover angles below $5^o$ and $10^o$ in the vertical and horizontal 
directions, respectively.

Starting from the interaction point, we find a silicon microvertex detector, 
a main tracker system, a Time-of-Flight (TOF) barrel and a Barrel DIRC.
To cover the forward angles a system of Drift Chambers (DC) 
will be added inside the TS and in the FS, as well.
The same thing for DIRC, where an endcap will be used.
The Muon detector will be embedded in the return yoke of the magnet, either 
to cover the azimuthal angle or to cover the forward angles.
In the FS different systems of DC, TOF, RICH, $\mu$ detectors and
hadron calorimeter will be used.
For the electromagnetic shower detection an EM calorimeter will be available 
in the TS and a Shashlyk calorimeter will be used in the FS.

Concerning the Hydrogen internal target two different systems are 
under development, one based on Pellet and one on Cluster-Jet.
Pellet target will guarantee to reach
the maximum luminosity, since an averaged thickness of
$10^{16}$~H/cm${^2}$ can be achieved. On the contrary Cluster-Jet target 
will assure to have a stable luminosity, since the density can be easily 
changed by means of the pressure.
Nuclear target, shaped in wires or foils will be also used for hypernuclear
physics and for the study of charm in the nuclear medium.

Main feature of the silicon microvertex detector are the following:
good vertex resolution, better than $100~\mu$m, to reconstruct 
displaced secondary vertices from charmed and strange hadron decays;
high granularity; low material budget and low power consumption.
The structure of the detector consists of four barrel layers
surrounding the interaction point and six disks in the forward region.
The two innermost barrel layers are made out of custom hybrid 
active pixel sensors~\cite{panda-pixel} of $100\times100\mu$m$^2$ size 
with an epitaxial sensor thickness less than $100\mu$m.
The four first disks in the forward direction, as well as an inner part
of the two outer disks, are also made out of hybrid pixel
modules to cope with the high particles flux.
For the readout system of the inner part a custom pixel chip 
(TOPIX~\cite{panda-topix}), based on CMOS 130~nm
technology is under study at Torino INFN, the layout with analog and 
digital parts fits 100$\%$ the pixel cell.
In the second prototype currently developed the time-over-threshold
has been implemented, in order to retain energy loss information.
Sufficient buffering to operate without trigger is forseen.
The sensor and the readout chip are bump-bonded together 
including the support structure and the cooling system.
The detector will have a radiation length X$_0$ about 1$\%$.
The reduced occupancy in the outer layers of the microvertex detector
allows to use double silicon microstrips, with a corresponding
reduction in the number of readout channels and material.
Both pitch size and thickness are 100~$\mu$m; such a system
will use a standard readout chip developed in other experiments.
The use of microstrip modules for the outer region allows to
reduce total material budget of the microvertex detector to 4-6$\%$ 
X$_0$. The total number of channel will be about 10$^7$.

For the tracker system two different options are under investigation.
One is based on Straw Tubes~\cite{panda_tracker} of Mylar,
10~mm radius and 1.5~m long. The inner and outer layers 
in the axial direction and the intermediate layers are skewed of a small 
angle up to $3^o$ in order to measure the z~coordinate.
It is a very light detector with a momentum resolution of 1.5$\%$.
The second option is a TPC with a multi-GEM system for avalanche
amplification (which allows to suppress the ion feedback).
Due to the high rate pf $\bar{p}$ annihilations ($10^7/s$)
the TPC has to operate continuously, therefore the gating technique 
cannot be applied.
The TPC is an ideal device for tracking due to low material budget,
a momentum resolution of 1$\%$ and a very good PID can be achieved,
provided the problems of the space charge build-up 
and the continous sampling have been solved.
Particles emitted at polar angles below $22^o$ will be tracked in the 
DC placed downstream the target.

A very good PID over a large range of energies is an essential 
requirement to fullfill the physics goals of PANDA.
Therefore, different processes will be used to assure redundancy:
$\check{C}$erenkov effect above 1~GeV/c, using different types of radiators;
energy loss below 1~GeV/c; TOF below 1~GeV/c; energy loss at low energy 
by means of microvertex detector; electromagnetic showers detection by means 
of EM calorimeters. 

In the TS the EM calorimeter~\cite{panda_emcal} 
will cover almost the full solid angle and will have a good efficiency 
and energy resolution up to 10~GeV and down to 10~MeV. 
It consists of one barrel and two endcaps, made of lead tungstate
PWO-II crystals of second generation, with enhanced light ouput. 
Large Area APD, which can be operated in 
presence of magnetic field, will be used for scintillating light
read-out, allowing a fast timing at a level of about 1~ns.
The PWO-II crystals will be operated at a temperature down to
T~=~-20$^o$C in order to increase the light yield by a further 
factor 3. 
An energy resolution of 2$\%$ at 1~GeV
can be achieved.
A Shashlyk calorimeter will be installed in the FS
beyond the dipole magnet, about 7~m from the interaction point.
Here a worse resolution of 4$\%$~+~const. at
1~GeV can be achieved at a moderate cost.

DAQ$\&$Trigger have to manage a high rate up to 20~MHz.
No hardware trigger will be implemented and a continuous sampling 
will be done. The detector Front Ends will be capable to make autonomous
data preprocessing. The data reduced in the preprocessing step 
will be marked by a precise time stamp and buffered for further processing.
The data will be sent to Powerful Compute Nodes, based on high density
FPGA preocessors.
A Configurable High Speed Network transports data through the levels.
The last stage of network can be more traditional and it will be attached 
to the online reconstruction farm, where the event is fully reconstructed 
to perform the final selection before the mass storage. 
Due to the high interaction rate a large 
bandwidth up to 200~GB/s is needed.

\end{document}